\documentclass[aps,prl,twocolumn,superscriptaddress,longbibliography]{revtex4-2}
\usepackage{graphicx}
\usepackage{dcolumn}
\usepackage{bm}
\usepackage{comment}
\usepackage{xcolor}
\usepackage{amsmath}

\begin{document}
\setcounter{secnumdepth}{3}
\preprint{APS/123-QED}

\title{Hybrid structure with a ferromagnetic film and an array of magnetic molecules for deep-nanoscale reprogrammable magnonics}
\author{Oleksandr Pastukh}
\thanks{These authors contributed equally to this work.}
\email{oleksandr.pastukh@ifj.edu.pl}
\affiliation{Institute of Nuclear Physics Polish Academy of Sciences, Krak\'ow, Poland}
\affiliation{Faculty of Physics and Astronomy, Adam Mickiewicz University, Pozna{\'n}, Poland}

\author{Piotr Graczyk}
\thanks{These authors contributed equally to this work.}
\email{graczyk@ifmpan.poznan.pl}
\affiliation{Institute of Molecular Physics, Polish Academy of Sciences, M. Smoluchowskiego 17, 60-179 Pozna\'{n}, Poland}

\author{Mateusz Zelent}
\thanks{These authors contributed equally to this work.}
\email{mateusz.zelent@amu.edu.pl}
\affiliation{Faculty of Physics and Astronomy, Adam Mickiewicz University, Pozna{\'n}, Poland}
\affiliation{Fachbereich Physik and Landesforschungszentrum OPTIMAS, Rheinland-Pfälzische Technische Universität
Kaiserslautern-Landau, 67663 Kaiserslautern, Germany}

 \author{{\L}ukasz Laskowski}
 \affiliation{Institute of Nuclear Physics Polish Academy of Sciences, Krak\'ow, Poland}
 
 \author{Maciej Krawczyk}
 \affiliation{Faculty of Physics and Astronomy, Adam Mickiewicz University, Pozna{\'n}, Poland}

\date{\today}
\begin{abstract}
Miniaturization is an essential element in the development of information processing technologies and is also one of the main determinants of the usability of the tested artificial neural networks. It is also a key element and one of the main challenges in the development of magnonic neuromorphic systems. In this work, we propose a new platform for the development of these new spin-wave-based technologies. Using micromagnetic simulations, we demonstrate that magnetic molecules regularly arranged on the surface of a thin ferromagnetic layer enable resonant coupling of propagating spin waves with the dynamics of the molecules' magnetic moments, opening a gap in the transmission spectrum up to 150 MHz. The gap, its width, and frequency can be controlled by an external magnetic field or the arrangement of molecules on the ferromagnetic surface. Furthermore, the antiferromagnetic arrangement of the magnetic moments of molecules or clusters of molecules allows for control of the gap's position and width. Thus, the proposed hybrid structure offers reprogrammability and miniaturization down to the deep nanoscale, operating frequencies in the range of several GHz, key properties for the implementation of artificial neural networks.

\end{abstract}


\maketitle

\section{\label{sec:level1}Introduction}

Magnon-based electronics has the great potential to revolutionize logic devices by overcoming the limitations of traditional CMOS technologies~\cite{wang2024nanoscale}, offering information transfer at high frequencies without charge current and thus with minimized energy dissipation \cite{kruglyak2010magnonics,chumak2015magnon}. However, precise control and manipulation of spin-wave (SW) propagation at the deep nanoscale is required to open up prominent prospects for wave-based data storage and logic applications of magnonics \cite{csaba2017perspectives,chappert2007emergence,khitun2010magnonic}.

The propagation of SWs over long distances, even centimeters, is enabled by magnetic materials exhibiting very low damping, such as yttrium iron garnet (Y$_3$Fe$_5$O$_{12}$, YIG) \cite{serga2010yig}. Various methods for preparing YIG ultra-thin films~\cite{Schmidt2020,Dubs2020} and, more recently, nanoscale waveguides~\cite{Mohseni2021,Nikolaev2023}, enable its use for low-loss SW transmission. Moreover, controlling different physical parameters, for instance geometry~\cite{Merbouche2021} or anisotropy~\cite{Medwal2021}, and effects, for instance an SW interaction with the magnetization texture \cite{Banerjee2017,YU20211,Szulc2022}, allows for tuning of SW propagation. Such control is particularly important for the successful development of SW-based devices and often relies on confining SWs within geometrically patterned waveguides \cite{wang2019spin} or within domain walls (DWs) \cite{han2019mutual}. Both of these approaches significantly affect the dispersion relation of SWs. However, both face limitations in logic and reprogrammable devices due to limited flexibility in controlling the propagation path and the mismatch in dynamic behavior between DWs and SWs. Furthermore, nanostructuring \cite{Nikolaev2023,levchenko2025} and doping \cite{Bensmann2025} ferromagnetic dielectrics reduces the propagation length of SWs \cite{Heinz2020,voronov2025}, which further forces the miniaturization of magnonic devices.

\begin{figure*}[!htp]
\includegraphics[width=\textwidth]{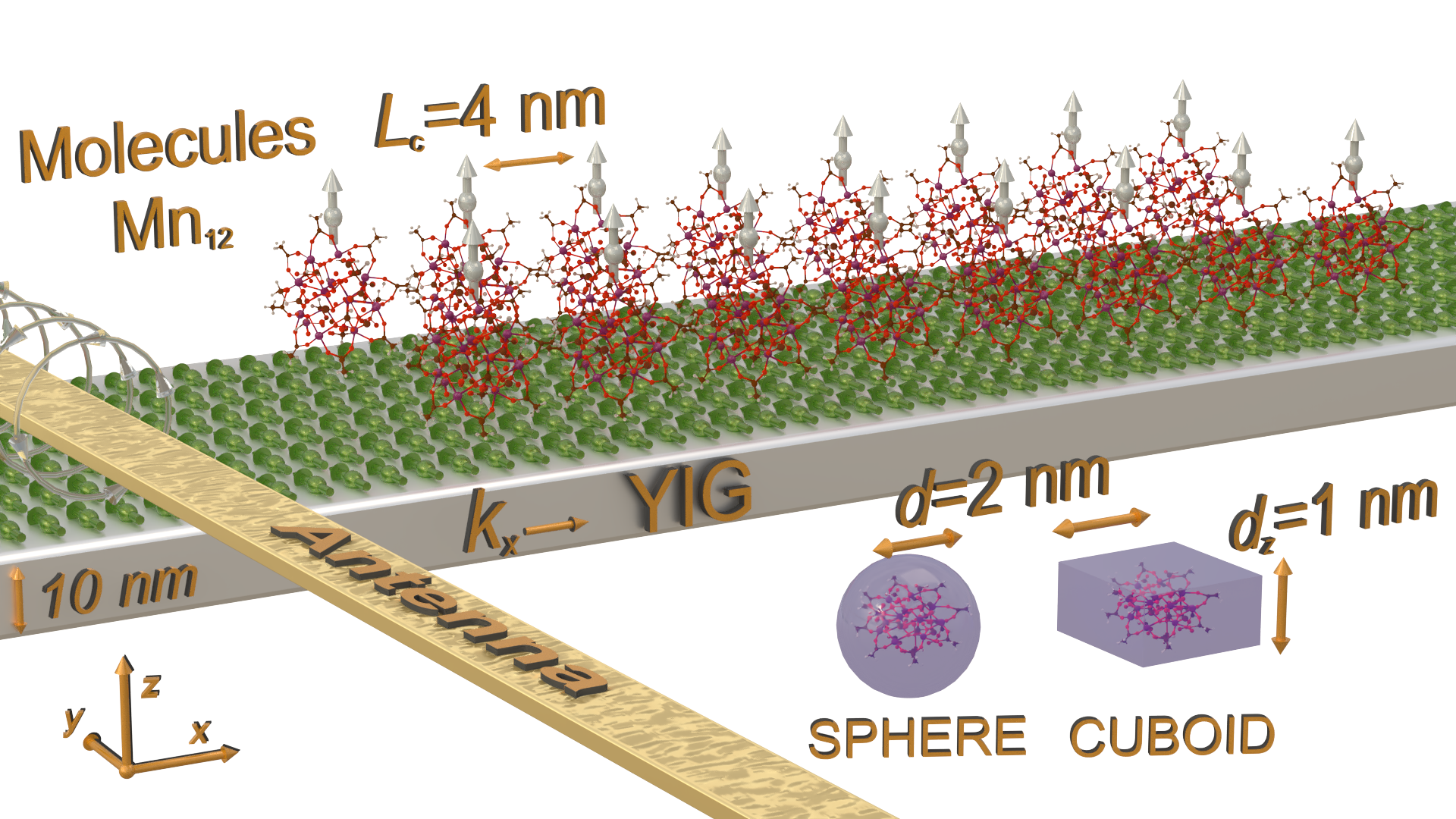}
\caption{\label{fig:system} Schematic illustration of the investigated hybrid magnonic system. The system consists of a 10 nm thick YIG film (shown in blue) with an array of Mn$_{12}$ single-molecule magnets (SMMs) deposited on its surface (shown as blue ellipses). The external magnetic field $H_{\text{ext}}$ is applied perpendicular to the spin wave propagation direction, establishing the Damon-Eshbach configuration. Spin waves are excited by a microwave antenna (shown as gold stripe) and propagate along the $x$-direction through the YIG film. The magnetic molecules are arranged in a square lattice with lattice constant $a$ and are positioned 1 nm above the YIG surface to avoid direct exchange coupling while maintaining dipolar interaction.}
\end{figure*}

Recent studies suggest that ferromagnetic nanostructures, e.g., nanostripes and nanodots, deposited on top of a ferromagnetic film may offer promising alternatives for the development of magnonic logic \cite{Mruczkiewicz2017,Graczyk2018old,Santos2023,Fripp2021,Szulc2025}. It was shown, for instance, that the dynamic dipolar coupling between a ferromagnetic metal nanostripe and a YIG film provides reconfigurable spin-wave transport and the system can operates as a Fabry-Pérot resonator \cite{qin2021nanoscale,Lutsenko2025}. 
The deposited array of Py nanodots can be used to transfer patterned magnetic information to the underlying YIG films by stray fields \cite{chaudhuri2022tuning}, significantly modifying transmission of SWs in homogeneous film~\cite{szulc2024}. Even more importantly, an array of ferromagnetic nanomagnets with perpendicular magnetic anisotropy, when placed on top of a ferromagnetic film, can be used as programming units of an artificial neural network. This is achieved through a specific landscape of static, stray magnetic fields generated by nanomagnets for SWs that propagate in the film \cite{papp2021nanoscale}.
The key advantage of SW-based computing lies in a high degree of interconnection between distinct spin-wave channels or ferromagnetic resonators~\cite{Kruglyak2021}, thereby allowing for the construction of two-dimensional networks. However, the experimental implementation encounters enormous difficulties, among others due to the strongly weakening spin-wave signal as a result of damping and scattering on the network of field barriers or dynamically coupled resonators.

Another step in this line of research is to achieve coupling and interconnection between isolated spins and SWs~\cite{Neuman2020} or spatially separated spins, as quantum objects, via propagating SWs. For instance, modern quantum information processing could potentially exploit spins on nitrogen-vacancy (NV) centers, which are known for their long spin-coherence times and facile quantum-state control \cite{childress2013diamond}. It has been demonstrated that SWs excited in a YIG layer can effectively interact with NV centers, enabling long-range coupling between spin qubits \cite{andrich2017long,fukami2021opportunities,kikuchi2017long}. Although such studies confirm the potential of SW and NV centers, practical quantum computation remains challenging due to the difficulty of engineering scalable long-range quantum gates. An alternative approach to enhancing quantum control and sensitivity in magnonic systems may lie in SW-mediated coupling of magnetic molecules.  This approach provides a larger spin value and allows for the use of chemical deposition methods.

Single-molecule magnets (SMMs) have long been considered promising candidates for future quantum computing and spintronic applications due to their excited electronic spin states and long quantum coherence times \cite{taran2021single,schlegel2008direct}. Moreover, these magnetic molecules exhibit high-spin ground states and magnetic bistability at low temperatures \cite{sessoli1993magnetic}, offering potentially valuable influence, via stray fields or dynamical coupling, on propagating SWs under molecules~\cite{Dey2025}.  Importantly, molecular magnets can be readily integrated into various substrate materials while preserving their unique magnetic properties \cite{gabarro2023magnetic}, thereby opening new avenues for their application in nanoelectronics. In our previous work, we demonstrated the technical feasibility of depositing Mn\textsubscript{12}-based SMMs onto surfaces \cite{laskowska2019magnetic,laskowski2019separation}, as well as controlling their distribution \cite{laskowska2019control,laskowska2020magnetic}. A similar strategy can be applied for deposition onto the surface of a monocrystalline YIG layer. 
However, two key questions remain unanswered. First, will there be coupling between the Mn$_{12}$ molecules and the SWs in the YIG substrate? If so, how strong will it be? Second, how do the density of SMMs and the relative orientation of their magnetic moments affect the propagation of SWs? Finding the answers to these questions will pave the way for the experimental realisation and use of hybrid --molecular–YIG systems in magnonic neural networks, as well as in quantum technologies. 


Here, we numerically investigate a hybrid magnonic platform comprising a thin YIG film decorated with an array of Mn\textsubscript{12}-based single-molecule magnets. Using micromagnetic simulations, we assess whether propagating spin waves couple to the molecular spins via dipolar interaction and identify the key parameters that govern this interaction. We focus on the roles of the external bias field, molecular density, propagation direction, and the relative orientation and clustering of neighboring molecular moments. Our findings point to a programmable, direction-sensitive coupling mechanism that can be exploited for molecule-based magnonics, magnonic neural networks, and quantum-magnonic interfaces.

\section{System description}

The proposed system is shown in Fig.~\ref{fig:system}. It consists of a 10-nm thick, 16 \textmu m-long YIG film (along the $x$-axis) with an array of Mn$_{12}$ molecules placed on its surface. The SWs are launched by the microstrip antenna on the left and propagate under the array; eventually, the signal is collected inductively by a similar microstrip on the right-hand side of the system. The bias magnetic field $H_0$ is applied in-plane to saturate magnetization in YIG. The propagating SWs generate a stray magnetic field that is sensed by the magnetic moment of the molecules, providing a mechanism for their dynamic coupling.

Calculations are performed in a micromagnetic framework by solving the Landau–Lifshitz–Gilbert (LLG) equation for the magnetization dynamics using the finite difference time domain (FTDT) method, and also, after linearization, the obtained eigenproblem is solved with the finite element method (FEM), see Section Methods for details. Since magnetization is a continuous function of position, the ferromagnetic nanoelement mimics the Mn\textsubscript{12}-type molecular magnet with volume and total magnetic moment roughly corresponding to the acetate derivative of this magnet [Mn$_{12}$O$_{12}$(CH$_3$(CH$_2$)$_{16}$CO$_2$)$_{11}$(CH$_3$ CO$_2$)$_5$(H$_2$O)$_4$]$\cdot$ 2CH$_3$COOH~\cite{burgert2007single}. 
Due to the lack of corresponding macroscopic parameters for Mn$_{12}$-based SMMs, these parameters were recalculated from the experimentally measured data, including the total magnetic moment of the molecule $\mu = 17\mu_\text{B}$ ($\mu_\text{B}$ is a Bohr magneton)~\cite{laskowska2019magnetic} and uniaxial anisotropy parameter $D = 0.66$ K \cite{verma2017ageing}. 
This high anisotropy, together with the exchange coupling between spins, allows the molecule to be treated as an artificial magnet~\cite{Eddins2014} and justifies the use of a micromagnetic approach.
Using the relations for the magnetization saturation $M_s = (\mu N_A)/V_{\text{mol}}$ and $K_u = D/V_{\text{mol}}$, where $V_{\text{mol}}$ is the molar volume of Mn$_{12}$ \cite{laskowska2019magnetic,da2008abrupt} the corresponding micromagnetic parameters are obtained: $M_s=41$ kA/m and $K_u=2.37$ kJ/m$^3$. The exchange stiffness constant $A = 20$ pJ/m, is estimated directly from the exchange integrals between the spins in Mn$_{12}$ molecule ($J_1$ = $J_2$ = 65 K and $J_3$ = $J_4$ = 10 K \cite{chaboussant2004exchange,honecker2005exchange}), by considering the molecule as a coarse-grained unit with an effective spin density and an average stiffness that reflects the contributions of all dominant exchange interactions. The damping value for the magnetic molecules is assumed to be the same as that of YIG.
For YIG film, the standard material parameters are taken: $M_\text{S} = 140$ kA/m, exchange stiffness constant $A = 3.5$ pJ/m, and the damping constant $\alpha_\text{YIG} = 2 \cdot  10^{-4}$.  

The assumed distribution of the molecules on the YIG layer can be physically reproduced by self-organization methods \cite{laskowski2019separation,laskowska2019control}. In this case, we assumed uniformly 2D-ordered molecular magnets on the YIG surface with a lattice constant of $L_\text{c}=4$~nm.
 In the case of surface-deposited SMMs, the magnetic molecules are often separated from the substrate \cite{laskowski2019separation}, which protects against exchange interactions between the molecule and the magnetic layer. Therefore, the molecules in the simulations were placed 1 nm above the YIG layer and the exchange interactions between the molecules and the YIG film were excluded.

\section{Results and discussion}
\subsection{Transmission spectra}
\begin{figure*}
\includegraphics[width=18cm]{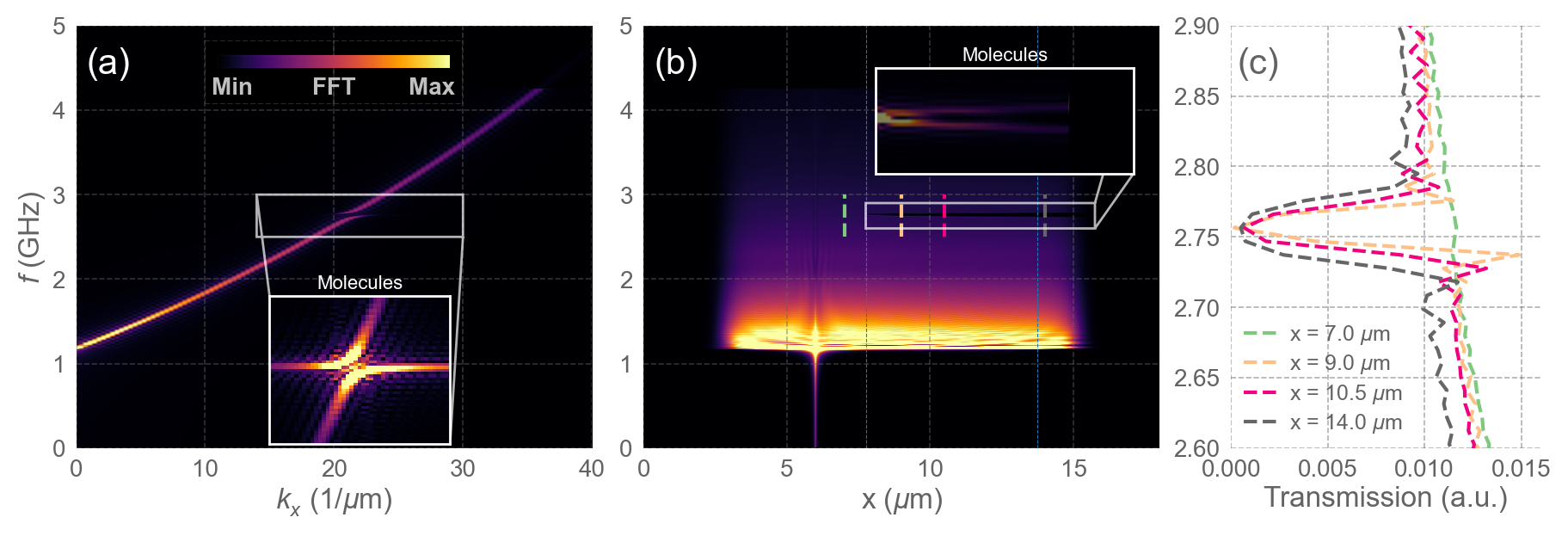}
\caption{\label{fig:mumax1} Spin wave transmission through the hybrid magnonic system obtained from time-domain simulations. (a) Dispersion relation of SWs in a YIG film decorated with a square array of Mn$_{12}$ molecules, showing the characteristic anti-crossing gap at 2.75 GHz. The inset displays the signal collected solely from the layer of molecules, confirming the coupling between the Mn$_{12}$ and YIG. (b) Evolution of SW transmission along the $x$-axis in the YIG layer. The inset shows the signal from the layer of molecules as a function of $x$, indicating energy transfer from the YIG to the molecules. (c) The transmission spectra at four selected positions: before ($x=7.0$ $\mu$m), within ($x=9.0$ and $10.5$ $\mu$m), and after ($x=14.0$ $\mu$m) the array of molecules. For all figures, SWs are excited at $x=6$ $\mu$m, and the molecular array occupies the region between $x=7.75$ $\mu$m and $x=13.75$ $\mu$m. The bias magnetic field $\mu_0 H_0 = 10$ mT is along the $y$ axis.}
\end{figure*}
 
The SWs are generated by the antenna placed in front of the molecule-covered area (located at $x = 6.0$ \textmu m, see Fig.~\ref{fig:system}).  First, we assume, the Damon–Eshbach (DE) configuration in which the external magnetic ﬁeld ($\mu_0 H_0 = 10$ mT) is directed along the $y$-axis, while the SW propagation is along the $x$-axis in the YIG ﬁlm plane. At $x=7.75$~\textmu m the SWs enter the area covered by the molecules.
The dispersion relation $f(k)$ ($f$ -- frequency, $k$ -- wavenumber) of SWs calculated from the YIG area is shown in Fig.~\ref{fig:mumax1}a. This spectrum demonstrates a typical parabolic magnon band originating from the ferromagnetic film, but with a gap of $\Delta f = 65$ MHz width (see Section Methods for the gap width calculation details), centered at 2.75~GHz (at $k_\Delta = 21.5$ \textmu m$^{-1}$). Taking the amplitude from the molecule layer yields the dispersion shown in the inset of  Fig.~\ref{fig:mumax1}a, which clearly indicates  the anti-crossing gap between the dispersion relation of SW propagating in YIG and the mode localized in the molecules at $f=2.75$~GHz. This anti-crossing gap demonstrates that propagating SWs are resonantly coupled to the mode of  molecules, and the gap width, $\Delta f$, is a direct measure of the coupling strength \cite{klingler2018spin,hu2023tunable}. 

 The transmission spectrum along the $x$-axis is shown in Fig.~\ref{fig:mumax1}b, along with its  cross-sections in Fig. \ref{fig:mumax1}c, which are taken at $x=7.0$ \textmu m (before the molecule array), $x=9.0$ \textmu m, $x=10.5$~\textmu m (inside the molecule array) and $x=14.0$ \textmu m (behind the molecule array).  Transmission through the molecule array is suppressed at their resonant frequency of $f=2.75$~GHz, starting from $x \approx 7.8$~\textmu m, i.e., the 15th row of the molecules. The dip in the transmission spectrum corresponds to the gap observed in the dispersion relation (Fig. \ref{fig:mumax1}a), with almost linear increase in the width with the propagation distance, from $\Delta f=$ 30 MHz at $x=10$ $\mu$m, to $\Delta f= 54$ MHz at the end of the array (an effect similar to one observed in magnonic crystals \cite{lee2009physical}). This widening of the dip is due to the low but non-zero group velocity of SW at the gap edges. It correlates also with the enhanced signal observed in the molecule array (see the inset in Fig. \ref{fig:mumax1}b), whose amplitude decreases with the distance from the array edge. Importantly, the depth of the dip, with transmission approaching 0, remains almost unchanged, starting from anti-crossing gap opening.

 Although the dynamic coupling between the propagating SWs and the magnetization dynamics of the SMMs has been demonstrated, many questions remain unanswered. In particular, is the assumed shape of the Mn molecules relevant to the coupling strength? Are the collective dynamics of the molecular array present and influential in the coupling strength? Can the magnitude and orientation of the bias magnetic field control the coupling strength? To answer these questions, we need to provide deep insight into the physics of the coupling between the molecule resonance mode and the propagating SWs in the YIG layer. To achieve this, we perform eigenfrequency simulations, and provide a qualitative analysis using coupled mode theory (CMT). Details of both approaches can be found in the Methods section.

\subsection{Mn$_{12}$--YIG coupling }
\begin{figure}[htp]
\includegraphics[width=\linewidth]{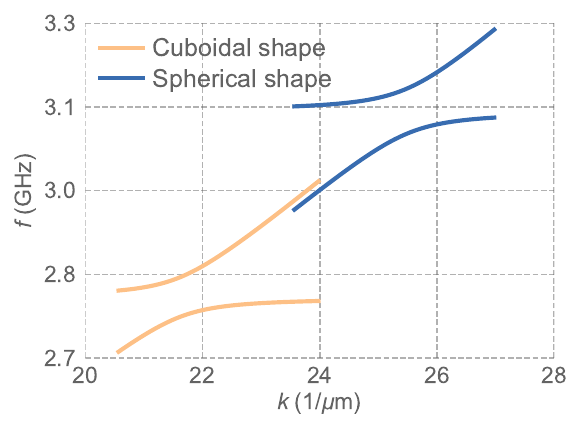}
\caption{\label{fig:square-circular} The dispersion relation of SWs in a hybrid structure around the anti-crossing between the DE SW mode and the molecule resonance for the molecules modeled as either cuboids (blue lines) or spheres (orange lines). Despite their different shapes, both geometries maintain the same total magnetic moment per molecule and the same distance between the molecular center and the YIG surface. The simulations were performed in the frequency domain using FEM. The bias magnetic field is $\mu_0 H_0=10$ mT.}
\end{figure}

  The FEM simulation results are consistent with the FDTD results, in terms of both $\Delta f$ and frequency positions, if the molecule has the form of cuboid (compare anti-crossing gap in the inset in Fig.~\ref{fig:mumax1}a and  Fig.~\ref{fig:square-circular}). When a spherical shape of a molecule is introduced to FEM simulations, the width of the anticrossing gap does not change (keeping the same value of the magnetic moment per molecule and the same distance between the centre of the molecule and the film surface as for a cuboid). However, the frequency position of the gap increases by 0.4 GHz, and the associated wavenumber increases from 21~$\mu$m$^{-1}$ to 25~$\mu$m$^{-1}$ for the spherical shape, as compared to the cuboid shape (see Fig.~\ref{fig:square-circular}). This indicates that the demagnetizing field related to the assumed shape of the molecule affects its resonance frequency, as it is for any size of the ferromagnetic element \cite{Joseph1965}.   Thus, the frequency/wave vector position of the anti-crossing gap should only be considered qualitatively, since the exact value of the demagnetizing field in the Mn$_{12}$ molecule is not known.
 Further FEM studies will be performed using a spherical shape to avoid the potential influence of the nonuniform demagnetizing fields.

\begin{figure*}
\includegraphics[width=\textwidth]{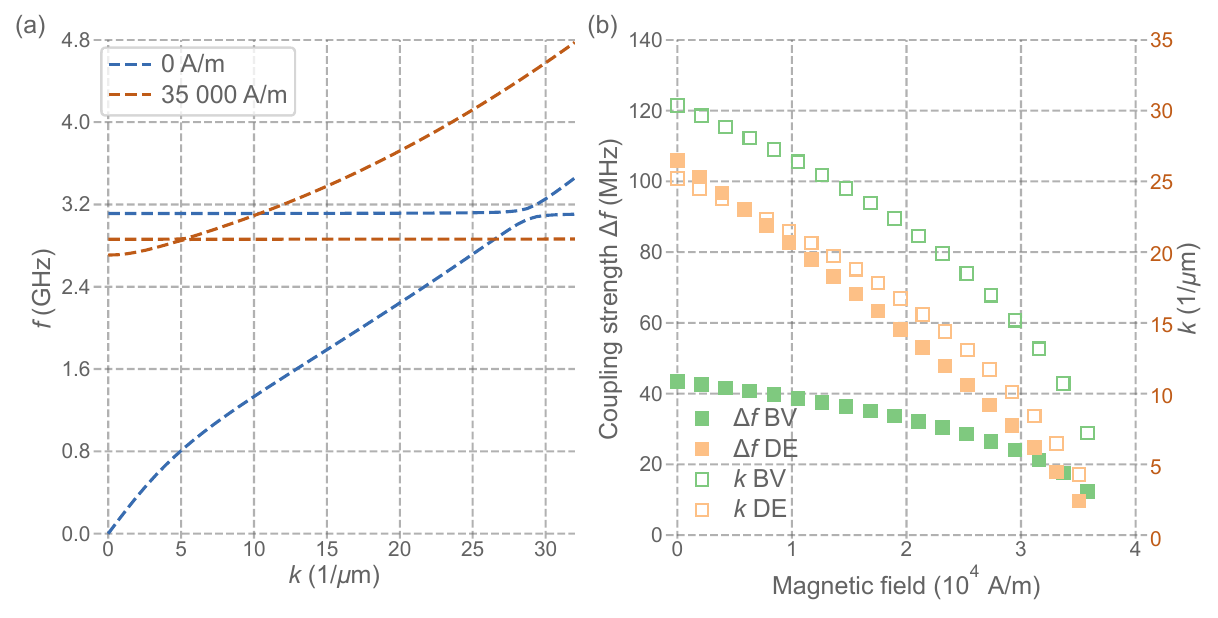}
\caption{\label{fig:H_1} (a) Dispersion relation of SWs in the hybrid system for the two values of the bias magnetic field in the DE geometry. (b) The coupling strength $\Delta f$ (full dots and the left-hand scale axis) between the resonances of the molecules and the propagating SWs in YIG in the DE geometry (filled red squares) and the BV geometry (filled green squares) as a function of the external magnetic field strength. The empty squares indicate the wavenumber at which the anti-crossing occurs (empty dots and the right-hand scale axis). 
}
\end{figure*}

In Fig.~\ref{fig:H_1}(a), we show the dispersion relations for two magnitudes of the bias magnetic field, $H_0=0$ and $35\times 10^3$~A/m,  in the DE configuration. As expected, the SW band of propagating SWs in YIG increases with the magnetic field (e.g., by 1.6~GHz at $k=1$~$\mu$m$^{-1}$ ). However, the frequency of the molecular resonance (the flat band) decreases slightly (by approximately 250 MHz) as the magnetic field increases. We attribute this frequency decrease to an increase in the tilt angle of the magnetization of the molecules away from the out-of-plane direction as the magnitude of the field increases. This tilt angle reaches almost 22$^\circ$ at $35\times 10^3$~A/m, resulting in a decreasing demagnetizing field along the $z$ axis. 
 The most important difference between the two spectra, however, is the difference in anti-crossing gap width, $\Delta f = 105$~MHz (opened at $k_\Delta=29$~\textmu m$^{-1}$) at $H_0=0$, and the gap width is reduced below 10~MHz at $H_0=35\times 10^3$~A/m (gap opened at $k_\Delta=6$~\textmu m$^{-1}$).
At first glance, rotating the magnetization of the molecules from the $z$ axis to the in-plane direction should strengthen the dynamical coupling between the magnetization dynamics of the molecules and the propagating SWs, because the dynamic component of the molecules' magnetization becomes more parallel to the out-of-plane stray field from the SWs. However, simulations show the opposite effect: the anti-crossing gap width decreases as the magnitude of the in-plane oriented magnetic field increases, as is further elaborated in Fig.~\ref{fig:H_1}(b) (see the full orange dots for the DE geometry). We found that this effect correlates with the anti-crossing gap shifting to lower wavenumbers, as shown by the empty orange squares and the vertical scale on the right in Fig.~\ref{fig:H_1}(b). This is because,  as the wavelength of the propagating SWs increases, the amplitude of the stray field decreases,  until it reaches 0 at $k\rightarrow \infty$.


This explains also a similar dependence between $\Delta f(H_0)$ and $k_\Delta(H_0)$ found in the backward volume (BV) geometry, i.e., when the magnetic field and magnetization in the YIG film are parallel to the wavevector, as shown in Fig.~\ref{fig:H_1}(b) with green full and empty dots. The anti-crossing gap width decreases from 45~MHz ($k_\Delta \approx 14$ \textmu m$^{-1}$) at 0 field to around 15~MHz ($k_\Delta \approx 6$ \textmu m$^{-1}$) at $H_0=35 \times 10^3$~A/m.  Nevertheless, the coupling strength between SWs and molecules in the BV geometry is over twice as weak as in the DE geometry at small bias fields ($\Delta f =45$ MHz vs. 105 MHz at $H_0=0$). As $H_0$ increases, the difference decreases until it disappears completely at $H_0=3.3 \times 10^3$ A/m. At this bias field, the wavenumber is 7 and 12~\textmu m$^{-1}$, respectively, for BV and DE. This indicates that the weaker stray dynamic magnetic field produced by SW in BV geometry is responsible for the weaker coupling. As the stray magnetic field in the DE geometry is strongly nonreciprocal and enhanced or suppressed on the opposite surfaces of the film, changing the propagation direction,  reversing the bias magnetic field, or placing molecules on the bottom surface of the YIG film will result in a significant decrease in coupling strength, e.g., in transmission spectra simulations the FWHM is reduced from 54 MHz (Fig.~\ref{fig:mumax1} for $x$ = 14 $\mu$m) to 32.5 MHz (see, supporting figures in the Supplementary Material). This demonstrates a chiral dynamical coupling between SWs and the magnetization dynamics of molecules. In fact, this is a property inherent to SWs in thin ferromagnetic films \cite{Kruglyak2021}.

\begin{figure}
\includegraphics[width=8.5cm]{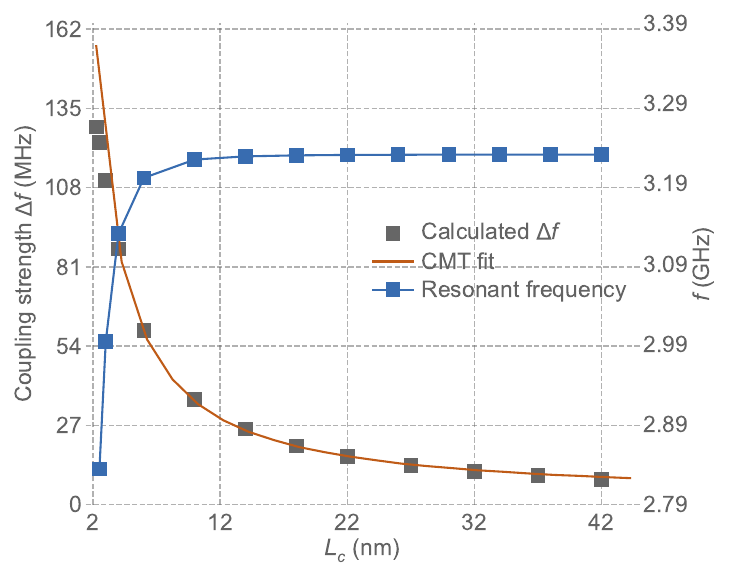}
\caption{\label{fig:coupling-strength} (a) The dependence of the resonance frequency of two molecules on the distance between them, showing the influence of direct dipolar interactions. (b) The coupling strength $\Delta f$ between the DE mode and the magnetic molecules as a function of intermolecular distance. Black points represent values extracted from numerical simulations and the red line shows the theoretical prediction from CMT. The bias field is set to zero. }
\end{figure}

An interesting question is whether and how direct dipolar interactions between molecules affect the strength of the dynamical coupling between the molecular dynamics and propagating SWs.  To elucidate this effect, we calculate the resonance frequency of the array of molecules at a zero bias field as a function of lattice constant, $L_\text{c}$. The results in Fig.~\ref{fig:coupling-strength}a (blue squares) show that the main resonance frequency decreases from approximately 3.22 GHz for well-separated molecules ($L_\text{c}=42$ nm) to slightly below 3.00 GHz at a 1 nm separation between molecules ($L_\text{c}=3$ nm). Given such a change in the resonance frequency, the wavenumber at which coupling with SW occurs changes by less than 1 \textmu m$^{-1}$. 
Interestingly, when molecules are deposited on the YIG film, the anti-crossing gap width increases from 10 to 130~MHz  as the lattice constant decreases from 42 to 3~nm, as shown by the black dots in Fig.~\ref{fig:coupling-strength}. This significant increase in $\Delta f$ can be attributed to the number of molecules within the distance covered by the half-wavelength of the propagating SWs, $\lambda/2$, where the molecules respond synchronously to the stray magnetic field driving them from the SWs. From Fig.~\ref{fig:H_1} we can estimate that $\lambda \approx 224$ nm at the anti-crossing gap ($k_\Delta=28$~\textmu m$^{-1}$), which does not change significantly with $L_\text{c}$. This means only five Mn$_{12}$ molecules in-row per wavelength at $L_\text{c}=3$ nm, and more than hundred molecules at $L_\text{c}=42$ nm.
In fact, an increase in coupling strength as the number of spins increases is a common feature, including for magnon-photon coupling involving molecules and microwave cavities~\cite{Eddins2014}.

The function $\Delta f(L_\text{c})$, obtained from CMT (Eq.~(\ref{Eq:kappa}), see also the Method section for details and the derivation) and neglecting any dynamic dipolar coupling between molecules, is plotted in Fig.~\ref{fig:coupling-strength} with a solid orange line. It accurately describes the gap width down to lattice constant $L_\text{c}=5$ nm (edge to edge separation of about 3 nm). At smaller separations, however, the model assuming non-interacting dipole moments breaks down, resulting in overestimation of the coupling strength. The CMT model predicts a value of 155 MHz, while from FEM simulations it is 130 MHz. This indicates that dynamical coupling between the molecules (directly via a stray field or indirectly via propagating SWs) plays a minor role in the coupling strength at separations larger than 3 nm, where the coupling is determined by the collective response of the array. However, it does influence the effective coupling between the molecular arrays and propagating SWs, decreasing it at high molecular densities.


\subsection{Programmability of the SW dynamics in Mn$_{12}$--YIG system}

\begin{figure}
\includegraphics[width=8.5cm]{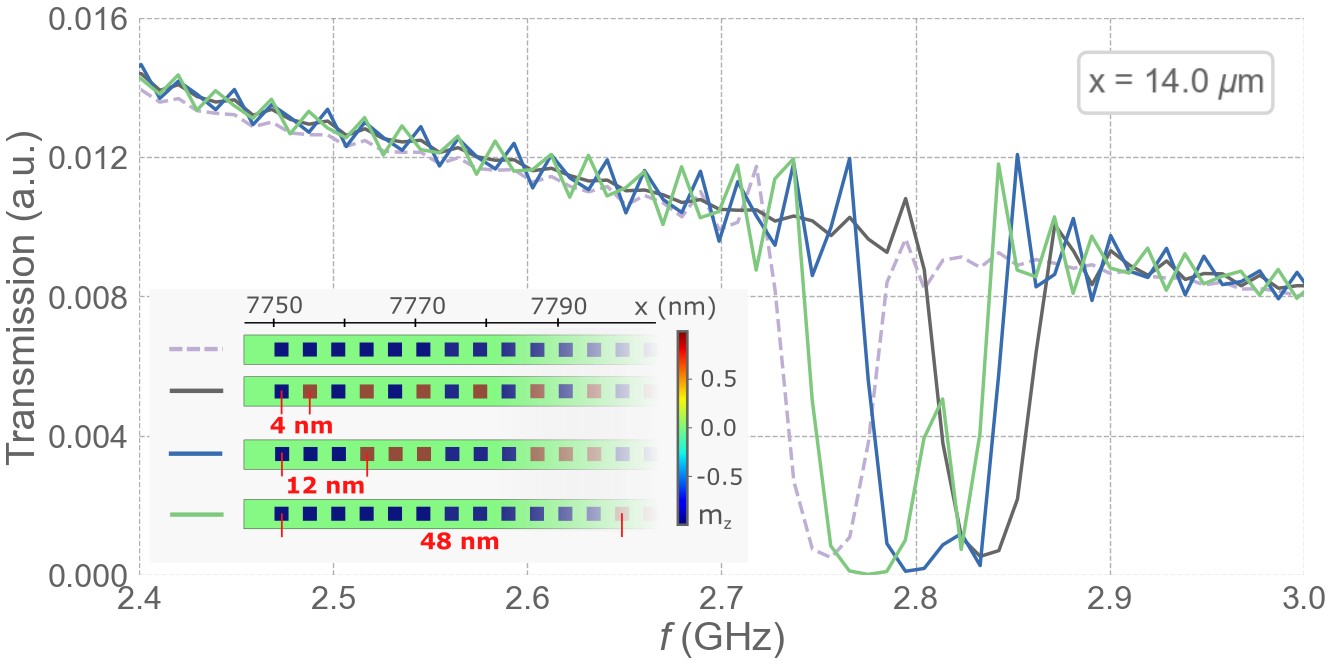}
\caption{\label{fig:AFM_2} Spin wave transmission observed through a hybrid system composed of a YIG film and a molecule array ($L_\text{c}=4$ nm) with magnetic moments in different configurations: ferromagnetic, antiferromagnetic (AFM), and AFM-ordered clusters composed of three molecules (with a cluster width of 12 nm) and twelve molecules (with a cluster width of 48 nm). The transmission spectra is obtained from FDTD simulations with spectra collected at a distance of 14 $\mu$m, i.e., in the YIG film behind the molecule array. The bias magnetic field is along the $y$-axis with a value of 10 mT.
}
\end{figure}
To determine whether the proposed system, which consists of molecules and a plain ferromagnetic film, is programmable and potentially useful for neuromorphic computations \cite{papp2021nanoscale}, we run simulations of the same system as before, but with all of the molecules' magnetic moments reversed. The results are the same as those presented in Fig.~\ref{fig:mumax1} (see the space-dependent transmission spectra and dispersion relation in the Supplementary Material). This indicates that the frequency of the anti-crossing gap and the coupling strength are independent of the orientation of the molecules' magnetic moments. However, when the molecules' moments are arranged antiferromagnetically (AFM) along the $x$ direction, the frequency anti-crossing gap shifts up by 90 MHz to 2.84 GHz, with approximately the same gap width as in a ferromagnetic order (see the dashed and black lines in Fig.~\ref{fig:AFM_2} for ferromagnetically and AFM ordered magnetic moments, respectively, and the figures in the Supplementary Material). This is a very important result. Since the observed frequency change is larger than the gap width ($\Delta f = 65$ MHz), the proposed system offers suitable programmability. Furthermore, the additional tunability of the anti-crossing gap position can be achieved through the formation of clusters (magnetic domains) of ferromagnetically ordered molecules.  Figure~\ref{fig:AFM_2} shows the transmission simulation results for clusters of three and twelve molecules that are ferromagnetically ordered but have opposite magnetic moment orientations in neighboring clusters.
Interestingly, the anti-crossing gap in the clustered system shifts frequency between the frequencies determined by ferromagnetically and AFM ordered magnetic moments and changes the gap width. For example, for three-molecule clusters, the frequency gap is 85 MHz. For twelve molecules' clusters we obtain the double anti-crossing separated by narrow (around 20 MHz) transmission band. These results clearly demonstrate the reprogrammability of the proposed hybrid structure. Due to its ability to control SW propagation, the structure is suitable for the realization of strongly miniaturized magnonic artificial neural networks.

\section{Practical realization and outlook}

The results presented in this work indicate that dipolarly coupled arrays of Mn$_{12}$ molecules placed on the surface of a thin YIG layer can interact strongly with propagating SWs and yield measurable anti-crossing gaps. From the perspective of experimental implementation of the proposed system, two elements are crucial: (i) stable immobilization and density control of Mn$_{12}$ on the YIG surface, and (ii) imparting a local magnetic orientation to the molecules (programmability).

(i) The issue of stable immobilization appears feasible via established techniques of chemical immobilization and spontaneous self-organization of molecules, as described previously \cite{ gomez2007advances,cornia2011chemical,adamek2023nanostructures}. Existing methods allow separations between molecules even on the order of a few nanometers and could provide statistical control over their surface distribution. Full ordering of deposited molecules, on the other side, can be achieved by employing e.g. stamp-assisted deposition of molecules from a solution \cite{cavallini2003multiple}, or using  a thin layer of porous silica with vertically aligned channels deposited on the YIG as a template for arranging the SMMs \cite{laskowski2019multi,laskowski2014spin}. Although such procedures has not yet been tested specifically in the Mn$_{12}$/YIG context, they can be regarded as a realistic and promising perspective.

(ii) The challenge of assigning a well-defined magnetic orientation is directly tied to the technological limitations above. Even though the microscopy techniques operating at low temperatures and under ultrahigh vacuum enable the probing of individual molecules immobilized on surfaces with spatial resolutions on the order of tens of nanometers \cite{gross2018atomic}, manipulation of magnetic properties of isolated magnetic molecules on surfaces is still challenging. Achieving full control over the magnetic state of each individual molecule without affecting neighboring units would require an array of independent magnetic force microscopy (MFM) tips spaced by specific nanometer distance. Thus, in practice, for a dense Mn$_{12}$ arrays, the only viable path is to impose a uniformly directed moment for all SMMs via a global magnetic field.
A more realistic scheme involves controlling the magnetic moments of entire Mn$_{12}$ clustered systems (e.g. composed of twelve or more molecules). Arranging independent MFM tips or other local field sources at scales of about 50-100 nm is already technologically attainable, though still challenging. For instance, MFM has been utilized to investigate local magnetic properties and to control the molecular orientation of arrays of TbPc$_2$ SMM dots deposited on modified SiO$_2$ films \cite{serri2017low}. Such an approach may pave the way for controlling the orientation of SMMs on the surface, at which molecular clusters exhibit distinct magnetization orientations, which opens up a possibility for realization of the reprogrammable YIG/Mn$_{12}$ structures presented in the study. Moreover, in this case the problem of nonuniform Mn$_{12}$ surface distribution is greatly mitigated -- statistically, the variations arising from a disordered molecular layout within each cluster average out, leading to comparable magnetic parameters across different areas.

Therefore, one can begin experimental realization with the magnetization of well-defined Mn$_{12}$ clusters on the surface, and gradually, as fabrication techniques improve, reduce the cluster size down to the single-molecule scale. This approach also allows identifying the threshold cluster size at which statistical fluctuations cease to significantly affect the local magnetic parameters and spin-wave propagation. Below that threshold one would adopt an ordered silica matrix to ensure full regularity in surface distribution.

This strategy simplifies fabrication and enhances robustness against errors in individual elements. That said, experimental realization demands careful investigation of several factors, notably the influence of inter-molecular distances and molecule height above the YIG surface on coupling strength, and the magnetic stability of Mn$_{12}$ under operational conditions. These issues stem directly from our simulations and should be the focus of further experimental follow-up.

\section{Conclusions}

In this study, we analyzed the possibility of coupling of individual magnetic molecules uniformly distributed on the surface of a thin YIG layer, with the SWs that propagate in that film. We demonstrate pronounced dynamical coupling of SWs with the molecule magnetic moment, which are modeled as magnetic nanostructures with properties corresponding to Mn$_{12}$ molecules. The energy of the SW excited in the layer is transferred to the molecules via the dipolar stray field, leading to the excitation of their spin oscillations. As a result, a resonance occurs near the intrinsic frequency of the magnetic molecules and wave vectors related to the SWs in YIG layer at this frequency. In the spin-wave dispersion, the coupling transfers into an anti-crossing frequency gap, which translates into the strong suppression of spin wave transmission under the molecule arrays.  

A key factor affecting the strength of the coupling between molecules and the YIG layer is the intermolecular distance. We show that by adjusting only the spacing between the molecular magnets—thereby modifying their mutual dipolar coupling—we can significantly influence spin-wave propagation in the YIG layer. For distances between Mn$_{12}$ molecules greater than 3 nm, dipole couplings between the magnetic molecules are negligible and have a minor effect on spin-wave propagation in YIG. In contrast, at shorter distances, these interactions must be taken into account as they significantly strengthen the resonance with the waveguide.  

Importantly, we demonstrate also a reprogrammability of the proposed system, which can be introduced by formation of the molecule clusters with antiferromagnetically oriented magnetic moments. This allows to tune the frequency and the width of the anti-crossing gap. A similar structure, but of much large size (over an order of magnitude larger), has been proposed to construct nanoscale neural networks based on nonlinear spin-wave interference \cite{papp2021nanoscale}. 
Our work explores the possibility of physically implementing such a circuit but with a much higher density of magnetic elements, and exploiting dynamical coupling between SWs and molecule dynamics.
We believe the presented results enable a feasible path toward true hardware implementations of artificial neural networks in nanoscale.


\section{Methods}
\subsection{Micromagnetic simulations}
Micromagnetic simulations are performed using an in-house version of the open-source, GPU-accelerated MuMax3~\cite{vansteenkiste2014design,Leliaert_2018} software, called AMUmax \cite{MathieuMoalicAMUmax}, which solves the LLG equation. In the effective magnetic field we include contributions from dipole interactions, exchange interactions, uniaxial anisotropy (only in molecules) and the external magnetic field. Due to the finite difference method used, each molecule is modeled as a cuboid with dimensions of $2\times 2 \times 1$~nm$^3$, which approximates the volume of the Mn$_{12}$ molecule \cite{garanin2008dipolar}. The nanoelements are arranged in a square array, with $L_\text{c}=4$ nm separation between the centres of the nearest neighbours.
To simplify the simulations, calculations were performed for a single molecule period along the $y$-direction with periodic boundary conditions (i.e., 256 repetitions in the $+y$ and $ -y$ directions). The total system size along the $x$ direction is 18 $\mu$m. The system is discretised into cells with dimensions of 1 × 1 × 1 nm$^3$ along the $x$, $y$ and $z$ axes, respectively.

\begin{figure}[h]
\includegraphics[width=8.5cm]{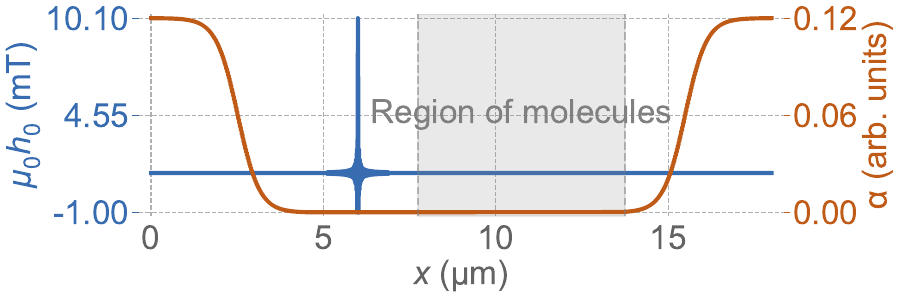}
\caption{
Spatial distribution of the external magnetic field amplitude 
$\mu_0 h_0$ (left axis) and the damping parameter $\alpha$ (right axis) 
along the sample. The shaded region marks the area containing molecular layers region of molecules.
}\label{fig:damping}
\end{figure}

 To excite the SWs (after initial relaxation of the system), an external microwave magnetic field is applied in the form of a sinc function in both the spatial and temporal domains, expressed as $h_{\text{rf}} = h_0 \text{sinc}[k_{\text{cut}} (x-x_0)]\text{sinc}[2\pi f_{\text{cut}} (t-t_0)]$, with a peak amplitude of $h_0=5$~mT appearing at time $t_0$ and position $x_0$. The cutoff frequency is $f_{\text{cut}}$ = 4.25 GHz and the cutoff wave vector $k_{\text{cut}}=25 \cdot 10^7$ m$^{-1}$. To avoid SW reflections from the edges of the system, absorbing boundary conditions were introduced along the $x$ axis. These were implemented by gradually increasing the Gilbert damping parameter near the edges of the YIG film. For this purpose, the waveguide layer was discretized along its length into 200 non-overlapping rectangular regions near the system edges, each spanning 50 nm. Across these regions, the damping value was smoothly varied from the minimum value $\alpha_{\text{YIG}} = 0.0002$ to the maximum value $\alpha_{\text{max}} = 0.125$, starting approximately 5 $\mu$m from each end of the YIG layer, according to the hyperbolic tangent function:
 \begin{equation*}
\alpha(x) = \alpha_{\text{YIG}} + (\alpha_{\text{max}} - \alpha_{\text{YIG}}) \left[\frac{1}{2} \tanh \left(-(x-x_0) \cdot s\right) + \frac{1}{2} \right],
\end{equation*}
\noindent where $x_0 = 100$ is the center point of the transition region. Here, $x \in [0,200]$ is an integer index corresponding to spatial discretization steps, and $s = 0.04$ is a scale factor that adjusts the smoothness of the damping transition. 
The resulting damping profile is shown in Fig. \ref{fig:damping}.

During the 300 ns of simulations, the time-resolved and space-dependent magnetization is recorded with a time interval of 100 ps. The transmission spectra plotted in Fig.~\ref{fig:mumax1}b-c are collected at the end of simulation time. The SW dispersion relation (Fig.~\ref{fig:mumax1}a) is obtained by performing a two-dimensional Fourier transform of the recorded magnetization into the frequency-wavenumber space with signal collected from the YIG film (main plot) and only from the molecules (the inset).  

To numerically evaluate the coupling strength between the YIG and molecular modes, the anti-crossing region of the dispersion relation (Fig.~\ref{fig:mumax1}a) was fitted using the analytical expression for the resonance angular frequencies of two coupled oscillators \cite{bhoi2019abnormal}:
\begin{flalign*}
f_{\text{YIG}(\text{Mn}_{12})} = \tfrac{1}{2}  \Bigl[
(&f^0_\text{YIG} + f^0_{\text{Mn}_{12}}) \\
&\quad \pm \sqrt{(f^0_\text{YIG} - f^0_{\text{Mn}_{12}})^2 + 4(\Delta f)^2}
\Bigr], 
\end{flalign*}
where $f_\text{YIG}$ and $f_{\text{Mn}_{12}}$ denote the hybridized resonance frequencies of YIG layer and molecules, respectively; $f^0_\text{YIG}$ and $f^0_{\text{Mn}_{12}}$ are their frequencies in the absence of coupling, and $\Delta f$ represents the coupling strength. The frequencies in the above equations are all taken at the wavenumber where the noninteracting dispersions cross.

The coupling strength obtained from the transmission spectra was evaluated 
by fitting the resonance dips (Fig.~\ref{fig:mumax1}c and Fig. \ref{fig:AFM_2}) with the Gaussian function and determining the corresponding 
full width at half maximum (FWHM).

\subsection{FEM simulations}

We perform simulations in the frequency domain using the FEM with COMSOL Multiphysics to analyze the detailed dispersion relation and the coupling between propagating SWs and magnetization oscillations in the molecules. In these simulations, we solve the Landau-Lifshitz-Gilbert equation together with Gauss's law for the scalar magnetic potential.
First, the system is relaxed to its equilibrium state in a time-dependent study using the full LLG equation with $\alpha  = 0.5$. Then, a linearized eigenfrequency analysis is performed.

For the FEM simulations, we select a sphere to avoid any shape anisotropy. The spherical molecules have a radius of $R_{\text{mol}}=9.85\text{ Å}$ and form a square lattice above the YIG layer, with a lattice constant of $a=2R_{\text{mol}}+2$ nm and an edge-to-edge distance of $5\text{ Å}$ from the YIG layer surface. The magnetisation of the molecule is $M_S = 41$ kA/m to maintain the same total magnetic moment as in FDTD simulations. The molecules' uniaxial anisotropy is set along the $z$ direction, with the same anisotropy constant as in FDTD. The simulated system is a unit cell with Bloch boundary conditions along the $x$ and $y$ axes. Along the $z$-axis, we assume a sufficiently large empty space to allow good convergence of the results at long-wavelength SWs.

\subsection{Coupled mode theory}
In order to  interpret the numerical results regarding the interaction between the molecules and the YIG layer, we employ the formalism of coupled mode theory (CMT)~\cite{Graczyk2017,Graczyk2018,Graczyk2018b}. The dynamic coupling between the molecule and the propagating SWs can be described by the respective stray fields generated by both. The molecule mode field is represented by the dynamic magnetic field $\vec{h}_\text{m}$ while the dynamic stray magnetic field of the DE mode in the YIG layer is represented by $\vec{h}_\text{l}$.
Then the strength of the coupling $\kappa$ is given by:
\begin{equation*}
\kappa = \frac{1}{V} \int \vec{h}_m \cdot \vec{h}_l dV,
\end{equation*}
where $V$ is the unit cell volume. The coupling strength $\kappa$ is related to the anti-crossing gap width $\Delta f$ by the relation $\kappa=\Delta f/2$.
Since the dynamic components of the layer are $h_{\text{l},x}$ and $h_{\text{l},z}$, while for the molecules that are $h_{\text{m},x}$ and $h_{\text{m},y}$  (we assume here that the molecules are magnetized along the $z$ axis, while YIG film is fully saturated along the $y$ axis), the modes couple only through the $x$-component of the magnetic field. Thus we introduce $h_\text{m}\equiv h_{\text{m},x}$, $h_l\equiv h_{\text{l},x}$. The fields $h_\text{l}, h_\text{m}$ are normalized by the factors $n_i$ such that
\begin{equation*}
\frac{n^2_i}{V} \int h_i h_i dV= 1,
\end{equation*}
where $i$ stands for $\text{l}$ and $\text{m}$, i.e., YIG film and molecule, respectively. 
 However, for the simplicity we can limit the integration to the plane at the center of the film ($z=d=6.5$ nm) in the case of $h_\text{l}$ and to the plane at fixed distance $z=1~\AA$ in the case of $h_\text{m}$. This is justified as long as we are interested in the qualitative description of the effect.  Therefore, the integration is done in the $xy$ plane:
\begin{equation}
\kappa = \kappa_0 \frac{1}{S} \int_{-a/2}^{a/2} \int_{-a/2}^{a/2} h_\text{m} h_\text{l} dx dy, 
\label{Eq:kappa}
\end{equation}
where $S=a^2$. With Eq.~(\ref{Eq:kappa}), we restored the quantitative picture by the fitting of the $\kappa$ with the constant $\kappa_0$ to the results of FEM simulations, ie., to the $\Delta f/2$. 

The magnetic field in the selected unit cell, we select the cell from the middle of the array, indicated by $\vec{r}_{00}$, has a contribution from the stray field of all other molecules in the system. To take this into account we introduce dipolar sums and assume that the field $\vec{h}_\text{m}(\vec{r}_{00})$ is given by the sum of the fields of the independent magnetic dipoles:
\begin{equation}
\vec{h}_\text{m}(\vec{r}_{00})=\cos{kx}\sum_{i=-N_x}^{N_x} \sum_{j=-N_y}^{N_y} \frac{3\hat{r}_{ij}(\vec{m}\cdot \hat{r}_{ij})-\vec{m}}{4\pi r_{ij}^3} 
, \label{Eq:dipolar_field}
\end{equation}
where $\vec{r}_{ij}=(x+ia/2)\hat{x}+(y+ja/2)\hat{y}+d\hat{z}$. The $x$ component of the magnetic field from the SW is approximated by $h_\text{l}=\cos{kx}$. The phase component $\cos{kx}$ in Eq.~(\ref{Eq:dipolar_field}) is crucial for nonzero coupling strength at $k\neq 0$, since the integral in Eq.~(\ref{Eq:kappa}) is zero for $k=0$  and $d>a$. Eq.~(\ref{Eq:kappa}) is solved numerically in MATLAB for $k=25$~\textmu m$^{-1}$ and finite number of dipoles $2N_x \times 2N_y$, where $N_i=\pi/kL_c$, and compared with the numerical results.

\section{Acknowledgment}
This work was supported by Polish National Science Centre projects 2020/37/B/ST3/03936, 2024/53/B/ST3/02188, and   2023/07/X/ST3/00677. The numerical simulations were performed at the Poznan Supercomputing and Networking Center (Grant No. PL0095-01 and PL0478-01). MZ acknowledges that this project has received funding from the European Union´s Framework Programme for Research and Innovation HORIZON-MSCA-2024-PF-01 under the Marie Sklodowska-Curie Grant Agreement Project 101208951 – CNMA.



\bibliography{manuscript-v1.bib}

@misc{MathieuMoalicAMUmax,
    title = {{AMUmax, Source available: https://zenodo.org/records/14203078. DOI: 10.5281/ZENODO.14203078}},
    author = {{Mathieu Moalic} and {Mateusz Zelent}},
    url = {https://zenodo.org/records/14203078 https://github.com/kkingstoun/amumax},
    doi = {10.5281/ZENODO.14203078}
}

@article{Graczyk2018old,
  title = {Magnonic band gap and mode hybridization in continuous permalloy films induced by vertical dynamic coupling with an array of permalloy ellipses},
  author = {Graczyk, Piotr and Krawczyk, Maciej and Dhuey, Scott and Yang, Wei-Gang and Schmidt, Holger and Gubbiotti, Gianluca},
  journal = {Phys. Rev. B},
  volume = {98},
  issue = {17},
  pages = {174420},
  numpages = {6},
  year = {2018},
  month = {Nov},
  publisher = {American Physical Society},
  doi = {10.1103/PhysRevB.98.174420}
}

@article{Joseph1965,
    author = {Joseph, R. I. and Schlömann, E.},
    title = {Demagnetizing Field in Nonellipsoidal Bodies},
    journal = {Journal of Applied Physics},
    volume = {36},
    number = {5},
    pages = {1579-1593},
    year = {1965},
    month = {05},
    doi = {10.1063/1.1703091}
}

@misc{levchenko2025,
      title={1D YIG hole-based magnonic nanocrystal}, 
      author={K. O. Levchenko and K. Davídkov{\'a} and R. O. Serha and M. Moalic and A. A. Voronov and C. Dubs and O. Surzhenko and M. Lindner and J. Panda and Q. Wang and O. Wojewoda and B. Heinz and M. Urbánek and M. Krawczyk and A. V. Chumak},
      year={2025},
      eprint={2506.10591},
      archivePrefix={arXiv},
      primaryClass={cond-mat.mes-hall},
      url={https://arxiv.org/abs/2506.10591}
}

@article{Santos2023,
author = {Santos, Obed Alves and van Wees, Bart J.},
title = {Magnon Confinement in an All-on-Chip {YIG} Cavity Resonator Using Hybrid {YIG/Py} Magnon Barriers},
journal = {Nano Letters},
volume = {23},
number = {20},
pages = {9303-9309},
year = {2023},
doi = {10.1021/acs.nanolett.3c02388}
}

@article{Szulc2025,
    author = {Szulc, Krzysztof and Zelent, Mateusz and Krawczyk, Maciej},
    title = {Multifunctional magnonic platform based on the interplay between spin-wave waveguide and nanodots with PMA and DMI},
    journal = {APL Materials},
    volume = {13},
    number = {9},
    pages = {091108},
    year = {2025},
    month = {09},
        issn = {2166-532X},
    doi = {10.1063/5.0277362}
}

@article{Fripp2021,
  title = {Spin-wave control using dark modes in chiral magnonic resonators},
  author = {Fripp, K. G. and Shytov, A. V. and Kruglyak, V. V.},
  journal = {Phys. Rev. B},
  volume = {104},
  issue = {5},
  pages = {054437},
  numpages = {10},
  year = {2021},
  month = {Aug},
  publisher = {American Physical Society},
  doi = {10.1103/PhysRevB.104.054437},
  url = {https://link.aps.org/doi/10.1103/PhysRevB.104.054437}
}

@article{Szulc2022,
author = {Szulc, Krzysztof and Tacchi, Silvia and Hierro-Rodríguez, Aurelio and Díaz, Javier and Gruszecki, Paweł and Graczyk, Piotr and Quir{\'o}s, Carlos and Mark{\'o}, Daniel and Martín, Jos{\'e} Ignacio and V{\'e}lez, María and Schmool, David S. and Carlotti, Giovanni and Krawczyk, Maciej and Álvarez-Prado, Luis Manuel},
title = {Reconfigurable Magnonic Crystals Based on Imprinted Magnetization Textures in Hard and Soft Dipolar-Coupled Bilayers},
journal = {ACS Nano},
volume = {16},
number = {9},
pages = {14168-14177},
year = {2022},
doi = {10.1021/acsnano.2c04256}

}

@article{Mruczkiewicz2017,
  title = {Spin-wave nonreciprocity and magnonic band structure in a thin permalloy film induced by dynamical coupling with an array of Ni stripes},
  author = {Mruczkiewicz, M. and Graczyk, P. and Lupo, P. and Adeyeye, A. and Gubbiotti, G. and Krawczyk, M.},
  journal = {Phys. Rev. B},
  volume = {96},
  issue = {10},
  pages = {104411},
  numpages = {7},
  year = {2017},
  month = {Sep},
  publisher = {American Physical Society},
  doi = {10.1103/PhysRevB.96.104411}
}

@article{Banerjee2017,
  title = {Magnonic band structure in a {Co/Pd} stripe domain system investigated by Brillouin light scattering and micromagnetic simulations},
  author = {Banerjee, Chandrima and Gruszecki, Pawel and Klos, Jaroslaw W. and Hellwig, Olav and Krawczyk, Maciej and Barman, Anjan},
  journal = {Phys. Rev. B},
  volume = {96},
  issue = {2},
  pages = {024421},
  numpages = {8},
  year = {2017},
  month = {Jul},
  publisher = {American Physical Society},
  doi = {10.1103/PhysRevB.96.024421}
}

@article{YU20211,
title = {Magnetic texture based magnonics},
journal = {Physics Reports},
volume = {905},
pages = {1-59},
year = {2021},
note = {Magnetic texture based magnonics},
issn = {0370-1573},
doi = {https://doi.org/10.1016/j.physrep.2020.12.004},
author = {Haiming Yu and Jiang Xiao and Helmut Schultheiss}
}

@article{Merbouche2021,
author = {Merbouche, Hugo and Collet, Martin and Evelt, Michael and Demidov, Vladislav E. and Prieto, Jos{\'e} Luis and Mu{\~n}oz, Manuel and Ben Youssef, Jamal and de Loubens, Gr{\'e}goire and Klein, Olivier and Xavier, St{\'e}phane and D’Allivy Kelly, Olivier and Bortolotti, Paolo and Cros, Vincent and Anane, Abdelmadjid and Demokritov, Sergej O.},
title = {Frequency Filtering with a Magnonic Crystal Based on Nanometer-Thick Yttrium Iron Garnet Films},
journal = {ACS Applied Nano Materials},
volume = {4},
number = {1},
pages = {121-128},
year = {2021},
doi = {10.1021/acsanm.0c02382}
}

@article{Nikolaev2023,
author = {Nikolaev, Kirill O. and Lake, Stephanie R. and Schmidt, Georg and Demokritov, Sergej O. and Demidov, Vladislav E.},
title = {Zero-Field Spin Waves in {YIG} Nanowaveguides},
journal = {Nano Letters},
volume = {23},
number = {18},
pages = {8719-8724},
year = {2023},
doi = {10.1021/acs.nanolett.3c02725}
}

@Article{Bensmann2025,
author={Bensmann, Jannis
and Schmidt, Robert
and Nikolaev, Kirill O.
and Raskhodchikov, Dimitri
and Choudhary, Shraddha
and Bhardwaj, Richa
and Taheriniya, Shabnam
and Varri, Akhil
and Niehues, Sven
and El Kadri, Ahmad
and Kern, Johannes
and Pernice, Wolfram H. P.
and Demokritov, Sergej O.
and Demidov, Vladislav E.
and Michaelis de Vasconcellos, Steffen
and Bratschitsch, Rudolf},
title={Dispersion-tunable low-loss implanted spin-wave waveguides for large magnonic networks},
journal={Nature Materials},
year={2025},
month={Jul},
day={09},
doi={10.1038/s41563-025-02282-y}
}

@article{Dubs2020,
  title = {Low damping and microstructural perfection of sub-40nm-thin yttrium iron garnet films grown by liquid phase epitaxy},
  author = {Dubs, Carsten and Surzhenko, Oleksii and Thomas, Ronny and Osten, Julia and Schneider, Tobias and Lenz, Kilian and Grenzer, J{\"o}rg and H{\"u}bner, Ren{\'e} and Wendler, Elke},
  journal = {Phys. Rev. Mater.},
  volume = {4},
  issue = {2},
  pages = {024416},
  numpages = {15},
  year = {2020},
  month = {Feb},
  publisher = {American Physical Society},
  doi = {10.1103/PhysRevMaterials.4.024416}}

@misc{voronov2025,
      title={Exchange spin-wave propagation in Ga:YIG nanowaveguides}, 
      author={Andrey A. Voronov and Khrystyna O. Levchenko and Roman Verba and Kristýna Davídková and Carsten Dubs and Michal Urb{\'a}nek and Qi Wang and Dieter Suess and Claas Abert and Andrii V. Chumak},
      year={2025},
      eprint={2509.05050},
      archivePrefix={arXiv},
      primaryClass={cond-mat.other},
      url={https://arxiv.org/abs/2509.05050}, 
}

@article{Heinz2020,
author = {Heinz, Bj{\"o}rn and Brächer, Thomas and Schneider, Michael and Wang, Qi and Lägel, Bert and Friedel, Anna M. and Breitbach, David and Steinert, Steffen and Meyer, Thomas and Kewenig, Martin and Dubs, Carsten and Pirro, Philipp and Chumak, Andrii V.},
title = {Propagation of Spin-Wave Packets in Individual Nanosized Yttrium Iron Garnet Magnonic Conduits},
journal = {Nano Letters},
volume = {20},
number = {6},
pages = {4220-4227},
year = {2020},
doi = {10.1021/acs.nanolett.0c00657}

}

@article{Schmidt2020,
author = {Schmidt, Georg and Hauser, Christoph and Trempler, Philip and Paleschke, Maximilian and Papaioannou, Evangelos Th.},
title = {Ultra Thin Films of Yttrium Iron Garnet with Very Low Damping: A Review},
journal = {physica status solidi (b)},
volume = {257},
number = {7},
pages = {1900644},
doi = {https://doi.org/10.1002/pssb.201900644},
year = {2020}
}

@article{Medwal2021,
    author = {Medwal, Rohit and Deka, Angshuman and Vas, Joseph Vimal and Duchamp, Martial and Asada, Hironori and Gupta, Surbhi and Fukuma, Yasuhiro and Rawat, Rajdeep Singh},
    title = {Facet controlled anisotropic magnons in {Y$_3$Fe$_5$O$_{12}$} thin films},
    journal = {Applied Physics Letters},
    volume = {119},
    number = {16},
    pages = {162403},
    year = {2021},
    month = {10},
    doi = {10.1063/5.0064653}
}

@article{Mohseni2021,
  title = {Controlling the Nonlinear Relaxation of Quantized Propagating Magnons in Nanodevices},
  author = {Mohseni, M. and Wang, Q. and Heinz, B. and Kewenig, M. and Schneider, M. and Kohl, F. and L{\"a}gel, B. and Dubs, C. and Chumak, A. V. and Pirro, P.},
  journal = {Phys. Rev. Lett.},
  volume = {126},
  issue = {9},
  pages = {097202},
  numpages = {6},
  year = {2021},
  month = {Mar},
  publisher = {American Physical Society},
  doi = {10.1103/PhysRevLett.126.097202}
}

@article{Lutsenko2025,
    author = {Lutsenko, Anton and Fripp, Kevin G. and Flajšman, Lukáš and Shytov, Andrey V. and Kruglyak, Volodymyr V. and van Dijken, Sebastiaan},
    title = {Magnonic Fabry–Pérot resonators as programmable phase shifters},
    journal = {Applied Physics Letters},
    volume = {126},
    number = {8},
    pages = {082406},
    year = {2025},
    month = {02},
    doi = {10.1063/5.0251358}
}

@misc{szulc2024,
      title={Reconfigurable spin-wave platform based on interplay between nanodots and waveguide in hybrid magnonic crystal}, 
      author={Krzysztof Szulc and Mateusz Zelent and Maciej Krawczyk},
      year={2024},
      eprint={2404.10493},
      archivePrefix={arXiv},
      primaryClass={cond-mat.mes-hall},
      url={https://arxiv.org/abs/2404.10493}, 
}

@article{Kruglyak2021,
    author = {Kruglyak, V. V.},
    title = {Chiral magnonic resonators: Rediscovering the basic magnetic chirality in magnonics},
    journal = {Applied Physics Letters},
    volume = {119},
    number = {20},
    pages = {200502},
    year = {2021},
    month = {11},
    doi = {10.1063/5.0068820}
}

@article{Dey2025,
author = {Dey, Sourav and Rivero-Carracedo, Gonzalo and Shumilin, Andrei and Gonzalez-Ballestero, Carlos and Baldoví, Jos{\'e} J.},
title = {Coupling Molecular Spin Qubits with 2D Magnets for Coherent Magnon Manipulation},
journal = {Nano Letters},
volume = {25},
number = {26},
pages = {10457-10464},
year = {2025},
doi = {10.1021/acs.nanolett.5c01937}
}

@article{Graczyk2017,
  title = {Coupled-mode theory for the interaction between acoustic waves and spin waves in magnonic-phononic crystals: Propagating magnetoelastic waves},
  author = {Graczyk, Piotr and Krawczyk, Maciej},
  journal = {Phys. Rev. B},
  volume = {96},
  issue = {2},
  pages = {024407},
  numpages = {10},
  year = {2017},
  month = {Jul},
  publisher = {American Physical Society},
  doi = {10.1103/PhysRevB.96.024407},
  url = {https://link.aps.org/doi/10.1103/PhysRevB.96.024407}
}

@article{Graczyk2018,
doi = {10.1088/1367-2630/aabb48},
url = {https://dx.doi.org/10.1088/1367-2630/aabb48},
year = {2018},
month = {may},
publisher = {IOP Publishing},
volume = {20},
number = {5},
pages = {053021},
author = {Graczyk, Piotr and Zelent, Mateusz and Krawczyk, Maciej},
title = {Co- and contra-directional vertical coupling between ferromagnetic layers with grating for short-wavelength spin wave generation},
journal = {New Journal of Physics}
}

@article{Graczyk2018b,
  title = {Magnonic band gap and mode hybridization in continuous permalloy films induced by vertical dynamic coupling with an array of permalloy ellipses},
  author = {Graczyk, Piotr and Krawczyk, Maciej and Dhuey, Scott and Yang, Wei-Gang and Schmidt, Holger and Gubbiotti, Gianluca},
  journal = {Phys. Rev. B},
  volume = {98},
  issue = {17},
  pages = {174420},
  numpages = {6},
  year = {2018},
  month = {Nov},
  publisher = {American Physical Society},
  doi = {10.1103/PhysRevB.98.174420},
  url = {https://link.aps.org/doi/10.1103/PhysRevB.98.174420}
}

@article{Leliaert_2018,
doi = {10.1088/1361-6463/aaab1c},
url = {https://dx.doi.org/10.1088/1361-6463/aaab1c},
year = {2018},
month = {feb},
publisher = {IOP Publishing},
volume = {51},
number = {12},
pages = {123002},
author = {Leliaert, J and Dvornik, M and Mulkers, J and De Clercq, J and Milosevis, M V and Van Waeyenberge, B},
title = {Fast micromagnetic simulations on {GPU}—recent advances made with {Mumax3}},
journal = {Journal of Physics D: Applied Physics}
}

@article{Eddins2014,
  title = {Collective Coupling of a Macroscopic Number of Single-Molecule Magnets with a Microwave Cavity Mode},
  author = {Eddins, A. W. and Beedle, C. C. and Hendrickson, D. N. and Friedman, Jonathan R.},
  journal = {Phys. Rev. Lett.},
  volume = {112},
  issue = {12},
  pages = {120501},
  numpages = {5},
  year = {2014},
  month = {Mar},
  publisher = {American Physical Society},
  doi = {10.1103/PhysRevLett.112.120501},
  url = {https://link.aps.org/doi/10.1103/PhysRevLett.112.120501}
}

@article{Neuman2020,
  title = {Nanomagnonic Cavities for Strong Spin-Magnon Coupling and Magnon-Mediated Spin-Spin Interactions},
  author = {Neuman, Tom\'a\ifmmode \check{s}\else \v{s}\fi{} and Wang, Derek S. and Narang, Prineha},
  journal = {Phys. Rev. Lett.},
  volume = {125},
  issue = {24},
  pages = {247702},
  numpages = {5},
  year = {2020},
  month = {Dec},
  publisher = {American Physical Society},
  doi = {10.1103/PhysRevLett.125.247702},
  url = {https://link.aps.org/doi/10.1103/PhysRevLett.125.247702}
}

@article{laskowski2019separation,
  title={The separation of the {Mn$_{12}$} single-molecule magnets onto spherical silica nanoparticles},
  author={Laskowski, Lukasz and Kityk, Iwan and Konieczny, Piotr and Pastukh, Oleksandr and Schabikowski, Mateusz and Laskowska, Magdalena},
  journal={Nanomaterials},
  volume={9},
  number={5},
  pages={764},
  year={2019},
  publisher={MDPI}
}

@article{laskowska2019magnetic,
  title={Magnetic behaviour of {Mn$_{12}$}-stearate single-molecule magnets immobilized inside {SBA-15} mesoporous silica matrix},
  author={Laskowska, Magdalena and Ba{\l}anda, Maria and Fitta, Magdalena and Dulski, Mateusz and Zubko, Maciej and Pawlik, Piotr and Laskowski, {\L}ukasz},
  journal={Journal of Magnetism and Magnetic Materials},
  volume={478},
  pages={20--27},
  year={2019},
  publisher={Elsevier}
}

@article{laskowska2019control,
  title={How to Control the Distribution of Anchored, {Mn$_{12}$--}Stearate, Single-Molecule Magnets},
  author={Laskowska, Magdalena and Pastukh, Oleksandr and Ku{\'z}ma, Dominika and Laskowski, {\L}ukasz},
  journal={Nanomaterials},
  volume={9},
  number={12},
  pages={1730},
  year={2019},
  publisher={MDPI}
}

@article{laskowska2020magnetic,
  title={Magnetic behaviour of {Mn$_{12}$}-stearate single-molecule magnets immobilized on the surface of 300 nm spherical silica nanoparticles},
  author={Laskowska, Magdalena and Pastukh, Oleksandr and Konieczny, Piotr and Dulski, Mateusz and Zalsi{\'n}ski, Marcin and Laskowski, Lukasz},
  journal={Materials},
  volume={13},
  number={11},
  pages={2624},
  year={2020},
  publisher={MDPI}
}

@article{garanin2008dipolar,
  title={Dipolar ordering and quantum dynamics of domain walls in {Mn}$_{12}$ acetate},
  author={Garanin, DA and Chudnovsky, EM},
  journal={Physical Review B},
  volume={78},
  number={17},
  pages={174425},
  year={2008},
  publisher={APS}
}

@article{vansteenkiste2014design,
  title={The design and verification of MuMax3},
  author={Vansteenkiste, Arne and Leliaert, Jonathan and Dvornik, Mykola and Helsen, Mathias and Garcia-Sanchez, Felipe and Van Waeyenberge, Bartel},
  journal={AIP advances},
  volume={4},
  number={10},
  year={2014},
  publisher={AIP Publishing}
}

@article{verma2017ageing,
  title={Ageing effects on the magnetic properties of {Mn}$_{12}$-based Acetate and Stearate SMMs},
  author={Verma, Apoorva and Verma, Shilpi and Singh, Priti and Gupta, Anurag},
  journal={Journal of Magnetism and Magnetic Materials},
  volume={439},
  pages={76--81},
  year={2017},
  publisher={Elsevier}
}

@article{klingler2018spin,
  title={Spin-torque excitation of perpendicular standing spin waves in coupled {YIG}/{Co} heterostructures},
  author={Klingler, Stefan and Amin, Vivek and Gepr{\"a}gs, Stephan and Ganzhorn, Kathrin and Maier-Flaig, Hannes and Althammer, Matthias and Huebl, Hans and Gross, Rudolf and McMichael, Robert D and Stiles, Mark D and others},
  journal={Physical review letters},
  volume={120},
  number={12},
  pages={127201},
  year={2018},
  publisher={APS}
}

@article{hu2023tunable,
  title={Tunable magnon-magnon coupling mediated by in-plane magnetic anisotropy in synthetic antiferromagnets},
  author={Hu, Bo and He, Wei},
  journal={Journal of Magnetism and Magnetic Materials},
  volume={565},
  pages={170283},
  year={2023},
  publisher={Elsevier}
}

@article{wang2024nanoscale,
  title={Nanoscale magnonic networks},
  author={Wang, Qi and Csaba, Gyorgy and Verba, Roman and Chumak, Andrii V and Pirro, Philipp},
  journal={Physical Review Applied},
  volume={21},
  number={4},
  pages={040503},
  year={2024},
  publisher={APS}
}

@article{kruglyak2010magnonics,
  title={Magnonics},
  author={Kruglyak, VV and Demokritov, SO and Grundler, D},
  journal={Journal of Physics D: Applied Physics},
  volume={43},
  number={26},
  pages={264001},
  year={2010},
  publisher={IOP Publishing}
}

@article{chumak2015magnon,
  title={Magnon spintronics},
  author={Chumak, Andrii V and Vasyuchka, Vitaliy I and Serga, Alexander A and Hillebrands, Burkard},
  journal={Nature physics},
  volume={11},
  number={6},
  pages={453--461},
  year={2015},
  publisher={Nature Publishing Group UK London}
}

@article{csaba2017perspectives,
  title={Perspectives of using spin waves for computing and signal processing},
  author={Csaba, Gy{\"o}rgy and Papp, {\'A}d{\'a}m and Porod, Wolfgang},
  journal={Physics Letters A},
  volume={381},
  number={17},
  pages={1471--1476},
  year={2017},
  publisher={Elsevier}
}

@article{chappert2007emergence,
  title={The emergence of spin electronics in data storage},
  author={Chappert, Claude and Fert, Albert and Van Dau, Fr{\'e}d{\'e}ric Nguyen},
  journal={Nature materials},
  volume={6},
  number={11},
  pages={813--823},
  year={2007},
  publisher={Nature Publishing Group UK London}
}

@article{khitun2010magnonic,
  title={Magnonic logic circuits},
  author={Khitun, Alexander and Bao, Mingqiang and Wang, Kang L},
  journal={Journal of Physics D: Applied Physics},
  volume={43},
  number={26},
  pages={264005},
  year={2010},
  publisher={IOP Publishing}
}

@article{serga2010yig,
  title={YIG magnonics},
  author={Serga, AA and Chumak, AV and Hillebrands, Burkard},
  journal={Journal of Physics D: Applied Physics},
  volume={43},
  number={26},
  pages={264002},
  year={2010},
  publisher={IOP Publishing}
}

@article{wang2019spin,
  title={Spin pinning and spin-wave dispersion in nanoscopic ferromagnetic waveguides},
  author={Wang, Q and Heinz, B and Verba, R and Kewenig, M and Pirro, P and Schneider, M and Meyer, T and L{\"a}gel, B and Dubs, C and Br{\"a}cher, T and others},
  journal={Physical review letters},
  volume={122},
  number={24},
  pages={247202},
  year={2019},
  publisher={APS}
}

@article{han2019mutual,
  title={Mutual control of coherent spin waves and magnetic domain walls in a magnonic device},
  author={Han, Jiahao and Zhang, Pengxiang and Hou, Justin T and Siddiqui, Saima A and Liu, Luqiao},
  journal={Science},
  volume={366},
  number={6469},
  pages={1121--1125},
  year={2019},
  publisher={American Association for the Advancement of Science}
}

@article{qin2021nanoscale,
  title={Nanoscale magnonic Fabry-P{\'e}rot resonator for low-loss spin-wave manipulation},
  author={Qin, Huajun and Holl{\"a}nder, Rasmus B and Flaj{\v{s}}man, Luk{\'a}{\v{s}} and Hermann, Felix and Dreyer, Rouven and Woltersdorf, Georg and van Dijken, Sebastiaan},
  journal={Nature communications},
  volume={12},
  number={1},
  pages={2293},
  year={2021},
  publisher={Nature Publishing Group UK London}
}

@article{chaudhuri2022tuning,
  title={Tuning spin wave modes in yttrium iron garnet films with stray fields},
  author={Chaudhuri, Ushnish and Singh, Navab and Mahendiran, R and Adeyeye, Adekunle O},
  journal={Nanoscale},
  volume={14},
  number={33},
  pages={12022--12029},
  year={2022},
  publisher={Royal Society of Chemistry}
}

@article{papp2021nanoscale,
  title={Nanoscale neural network using non-linear spin-wave interference},
  author={Papp, {\'A}d{\'a}m and Porod, Wolfgang and Csaba, Gyorgy},
  journal={Nature communications},
  volume={12},
  number={1},
  pages={6422},
  year={2021},
  publisher={Nature Publishing Group UK London}
}

@article{childress2013diamond,
  title={Diamond NV centers for quantum computing and quantum networks},
  author={Childress, Lilian and Hanson, Ronald},
  journal={MRS bulletin},
  volume={38},
  number={2},
  pages={134--138},
  year={2013},
  publisher={Cambridge University Press}
}

@article{andrich2017long,
  title={Long-range spin wave mediated control of defect qubits in nanodiamonds},
  author={Andrich, Paolo and de las Casas, Charles F and Liu, Xiaoying and Bretscher, Hope L and Berman, Jonson R and Heremans, F Joseph and Nealey, Paul F and Awschalom, David D},
  journal={npj Quantum Information},
  volume={3},
  number={1},
  pages={28},
  year={2017},
  publisher={Nature Publishing Group UK London}
}

@article{fukami2021opportunities,
  title={Opportunities for long-range magnon-mediated entanglement of spin qubits via on-and off-resonant coupling},
  author={Fukami, Masaya and Candido, Denis R and Awschalom, David D and Flatt{\'e}, Michael E},
  journal={PRX quantum},
  volume={2},
  number={4},
  pages={040314},
  year={2021},
  publisher={APS}
}

@article{kikuchi2017long,
  title={Long-distance excitation of nitrogen-vacancy centers in diamond via surface spin waves},
  author={Kikuchi, Daisuke and Prananto, Dwi and Hayashi, Kunitaka and Laraoui, Abdelghani and Mizuochi, Norikazu and Hatano, Mutsuko and Saitoh, Eiji and Kim, Yousoo and Meriles, Carlos A and An, Toshu},
  journal={Applied Physics Express},
  volume={10},
  number={10},
  pages={103004},
  year={2017},
  publisher={IOP Publishing}
}

@incollection{taran2021single,
  title={Single-Molecule Magnets and Molecular Quantum Spintronics},
  author={Taran, Gheorghe and Bonet, Edgar and Wernsdorfer, Wolfgang},
  booktitle={Handbook of Magnetism and Magnetic Materials},
  pages={979--1009},
  year={2021},
  publisher={Springer}
}

@article{schlegel2008direct,
  title={Direct observation of quantum coherence in single-molecule magnets},
  author={Schlegel, C and Van Slageren, J and Manoli, M and Brechin, EK and Dressel, M},
  journal={Physical review letters},
  volume={101},
  number={14},
  pages={147203},
  year={2008},
  publisher={APS}
}

@article{sessoli1993magnetic,
  title={Magnetic bistability in a metal-ion cluster},
  author={Sessoli, Roberta and Gatteschi, Dante and Caneschi, Andrea and Novak, MA},
  journal={Nature},
  volume={365},
  number={6442},
  pages={141--143},
  year={1993},
  publisher={Nature Publishing Group UK London}
}

@article{gabarro2023magnetic,
  title={Magnetic molecules on surfaces: SMMs and beyond},
  author={Gabarr{\'o}-Riera, Guillem and Arom{\'\i}, Guillem and Sa{\~n}udo, E Carolina},
  journal={Coordination Chemistry Reviews},
  volume={475},
  pages={214858},
  year={2023},
  publisher={Elsevier}
}

@article{burgert2007single,
  title={Single-molecule magnets: a new approach to investigate the electronic structure of {Mn$_{12}$} molecules by scanning tunneling spectroscopy},
  author={Burgert, Michael and Voss, S{\"o}nke and Herr, Simon and Fonin, Mikhail and Groth, Ulrich and R{\"u}diger, Ulrich},
  journal={Journal of the American Chemical Society},
  volume={129},
  number={46},
  pages={14362--14366},
  year={2007},
  publisher={ACS Publications}
}

@book{da2008abrupt,
  title={Abrupt Changes in the Tunneling Levels for {Mn$_{12}$-tBuAc}},
  author={da Silva Neto, Eduardo H},
  year={2008},
  publisher={Amherst College}
}

@article{laskowski2019multi,
  title={Multi-step functionalization procedure for fabrication of vertically aligned mesoporous silica thin films with metal-containing molecules localized at the pores bottom},
  author={Laskowski, {\L}ukasz and Laskowska, Magdalena and Dulski, Mateusz and Zubko, Maciej and Jelonkiewicz, Jerzy and Perzanowski, Marcin and Vila, Neus and Walcarius, Alain},
  journal={Microporous and Mesoporous Materials},
  volume={274},
  pages={356--362},
  year={2019},
  publisher={Elsevier}
}

@article{honecker2005exchange,
  title={Exchange constants and spin dynamics in {Mn$_{12}$}-acetate},
  author={Honecker, A and Fukushima, N and Normand, Bruce and Chaboussant, G and G{\"u}del, H-U},
  journal={Journal of magnetism and magnetic materials},
  volume={290},
  pages={966--969},
  year={2005},
  publisher={Elsevier}
}

@article{chaboussant2004exchange,
  title={Exchange interactions and high-energy spin states in Mn 12-acetate},
  author={Chaboussant, G and Sieber, A and Ochsenbein, S and G{\"u}del, H-U and Murrie, M and Honecker, A and Fukushima, N and Normand, B},
  journal={Physical Review B—Condensed Matter and Materials Physics},
  volume={70},
  number={10},
  pages={104422},
  year={2004},
  publisher={APS}
}

@article{bhoi2019abnormal,
  title={Abnormal anticrossing effect in photon-magnon coupling},
  author={Bhoi, Biswanath and Kim, Bosung and Jang, Seung-Hun and Kim, Junhoe and Yang, Jaehak and Cho, Young-Jun and Kim, Sang-Koog},
  journal={Physical Review B},
  volume={99},
  number={13},
  pages={134426},
  year={2019},
  publisher={APS}
}

@inproceedings{laskowski2014spin,
  title={Spin-glass implementation of a hopfield neural structure},
  author={Laskowski, {\L}ukasz and Laskowska, Magdalena and Jelonkiewicz, Jerzy and Boullanger, Arnaud},
  booktitle={International Conference on Artificial Intelligence and Soft Computing},
  pages={89--96},
  year={2014},
  organization={Springer}
}

@article{gomez2007advances,
  title={Advances on the nanostructuration of magnetic molecules on surfaces: the case of single-molecule magnets (SMM)},
  author={G{\'o}mez-Segura, Jordi and Veciana, Jaume and Ruiz-Molina, Daniel},
  journal={Chemical Communications},
  number={36},
  pages={3699--3707},
  year={2007},
  publisher={Royal Society of Chemistry}
}

@article{cornia2011chemical,
  title={Chemical strategies and characterization tools for the organization of single molecule magnets on surfaces},
  author={Cornia, Andrea and Mannini, Matteo and Sainctavit, Philippe and Sessoli, Roberta},
  journal={Chemical Society Reviews},
  volume={40},
  number={6},
  pages={3076--3091},
  year={2011},
  publisher={The Royal Society of Chemistry}
}

@article{adamek2023nanostructures,
  title={Nanostructures as the substrate for single-molecule magnet deposition},
  author={Adamek, Micha{\l} and Pastukh, Oleksandr and Laskowska, Magdalena and Karczmarska, Agnieszka and Laskowski, {\L}ukasz},
  journal={International Journal of Molecular Sciences},
  volume={25},
  number={1},
  pages={52},
  year={2023},
  publisher={MDPI}
}

@article{cavallini2003multiple,
  title={Multiple length scale patterning of single-molecule magnets},
  author={Cavallini, Massimiliano and Biscarini, Fabio and Gomez-Segura, Jordi and Ruiz, Daniel and Veciana, Jaume},
  journal={Nano Letters},
  volume={3},
  number={11},
  pages={1527--1530},
  year={2003},
  publisher={ACS Publications}
}

@article{gross2018atomic,
  title={Atomic force microscopy for molecular structure elucidation},
  author={Gross, Leo and Schuler, Bruno and Pavli{\v{c}}ek, Niko and Fatayer, Shadi and Majzik, Zsolt and Moll, Nikolaj and Pe{\~n}a, Diego and Meyer, Gerhard},
  journal={Angewandte Chemie International Edition},
  volume={57},
  number={15},
  pages={3888--3908},
  year={2018},
  publisher={Wiley Online Library}
}

@article{serri2017low,
  title={Low-temperature magnetic force microscopy on single molecule magnet-based microarrays},
  author={Serri, Michele and Mannini, Matteo and Poggini, Lorenzo and V{\'e}lez-Fort, Emilio and Cortigiani, Brunetto and Sainctavit, Philippe and Rovai, Donella and Caneschi, Andrea and Sessoli, Roberta},
  journal={Nano letters},
  volume={17},
  number={3},
  pages={1899--1905},
  year={2017},
  publisher={ACS Publications}
}

@article{lee2009physical,
  title={Physical Origin and Generic Control of Magnonic Band Gaps of Dipole-Exchange Spin Waves<? format?> in Width-Modulated Nanostrip Waveguides},
  author={Lee, Ki-Suk and Han, Dong-Soo and Kim, Sang-Koog},
  journal={Physical review letters},
  volume={102},
  number={12},
  pages={127202},
  year={2009},
  publisher={APS}
}

\end{document}